# Reconfigurable quantum photonics with on–chip detectors

Samuel Gyger,[*,†] Julien Zichi,[†] Lucas Schweickert,[†] Ali W. Elshaari,[†] Stephan Steinhauer,[†] Saimon F. Covre da Silva,[‡] Armando Rastelli,[‡] Val Zwiller,[†] Klaus D. Jöns,[†] and Carlos Errando-Herranz[*,†]

[†]*Quantum Nanophotonics, KTH Royal Institute of Technology, Roslagstullsbacken 21, 10691 Stockholm, Sweden*

[‡]*Institute of Semiconductor and Solid State Physics, Johannes Kepler Universität, Linz, Linz Institute of Technology, Altenbergerstraße 69, 4040 Linz, Austria*

E-mail: gyger@kth.se; carloseh@kth.se

## Abstract

Integrated quantum photonics offers a promising path to scale up quantum optics experiments by miniaturizing and stabilizing complex laboratory setups. Central elements of quantum integrated photonics are quantum emitters, memories, detectors, and reconfigurable photonic circuits. In particular, integrated detectors not only offer optical readout but, when interfaced with reconfigurable circuits, allow feedback and adaptive control, crucial for deterministic quantum teleportation, training of neural networks, and stabilization of complex circuits.

However, the heat generated by thermally reconfigurable photonics is incompatible with heat–sensitive superconducting single–photon detectors, and thus their on–chip co–integration remains elusive. Here we show low–power microelectromechanical reconfiguration of integrated photonic circuits interfaced with superconducting single–photon detectors on the same chip. We demonstrate three key functionalities for photonic quantum technologies: 28 dB high–extinction routing of classical and quantum light, 90 dB high–dynamic range single–photon detection, and stabilization of optical excitation over 12 dB power variation. Our platform enables heat–load free reconfigurable linear optics and adaptive control, critical for quantum state preparation and quantum logic in large–scale quantum photonics applications.

# Introduction

Optical quantum technologies are crucial to materialize the promises of quantum communication,[1] quantum computing,[2] and quantum simulation.[3] These applications require a leap in system complexity, only achievable via the miniaturization and stability provided by large–scale photonic integrated circuits (PICs).[4]

A quantum PIC is formed by a set of building blocks such as single–photon sources, quantum memories, reconfigurable photonic circuits, and detectors.[5–7] Reconfigurable photonic circuits not only provide the link between the other building blocks, but also enable the linear optic operations required for quantum state preparation and quantum logic.[8] In particular, combining reconfigurable photonics with detectors is central for on–chip single–photon detection and to enable feedback and adaptive control. Feedback is essential for quantum communication and computation protocols based on deterministic teleportation,[9] for self–configuration of arbitrary linear optics,[10] and for monitoring and stabilization of power, phase, and polarization. Elements addressing these functions often outnumber any other device in proposed protocols and experimental setups, and thus their on–chip integration is a central challenge, often overlooked, towards the upscaling of classical and quantum optics.[11,12] This requires reconfigurable elements with low optical loss, a small footprint, and low electrical power consumption for cryogenic compatibility. Traditional PIC reconfiguration based on thermo-optic,[13] carrier dispersion,[14] and electro-optic $\chi^{(2)}$ effects[15] suffers from high power consumption, high optical loss, and large footprint respectively. A promising cryogenic compatible reconfiguration method is microelectromechanical (MEMS) actuation, which combines low power consumption, low optical loss, and small footprint.[16] However, to date, there has been no demonstration of the compatibility of reconfigurable photonics with single–photon detectors in the same quantum PIC.[17]

Here, we integrate MEMS PIC reconfiguration with superconducting single–photon detection on the same chip, and show three crucial components of quantum optics experiments. We demonstrate reconfigurable routing of classical light and single–photons, high–dynamic range detection of optical excitation powers and single–photons, and power stabilization of optical excitation using a feedback loop.

# Results

## Waveguide–coupled single–photon detectors

Superconducting nanowire single-photon detectors (SNSPDs) provide broadband detection of single–photons with high detection efficiency, high signal-to-noise ratio, fast recovery time, and low timing uncertainty.[18] Their compatibility with mature PIC material platforms[19] and their excellent on–waveguide performance makes them outstanding integrated single-photon detectors.[7]

In this work we fabricate hairpin[20] SNSPDs from a 9 nm thin NbTiN film[21] on top of $Si_3N_4$ waveguides (see Fig. 1a). The 65 nm wide and 40 µm long photo-sensitive part of the wire is connected in series to a lumped–element inductor that prevents latching.[22] The detectors exhibit a saturated detection regime at 795 nm wavelength, revealed by the sigmoidal shape of the detection efficiency versus the bias current (Fig. 1c), which indicates unity internal quantum efficiency.[23] The two detectors feature critical currents of 15.8 µA (detector A) and 5.9 µA (detector B), and different on-chip detection efficiencies,[24] which we attribute to film heterogeneity and fabrication variations leading to localized constrictions.[25] The devices show timing jitters of $121.0 \pm 1.9$ ps and $253.0 \pm 1.4$ ps with room temperature amplification, and reset times with an exponential decay of $4.73 \pm 0.03\,\text{ns}$ and $4.71 \pm 0.01\,\text{ns}$ (detectors A and B respectively, see Supplementary).

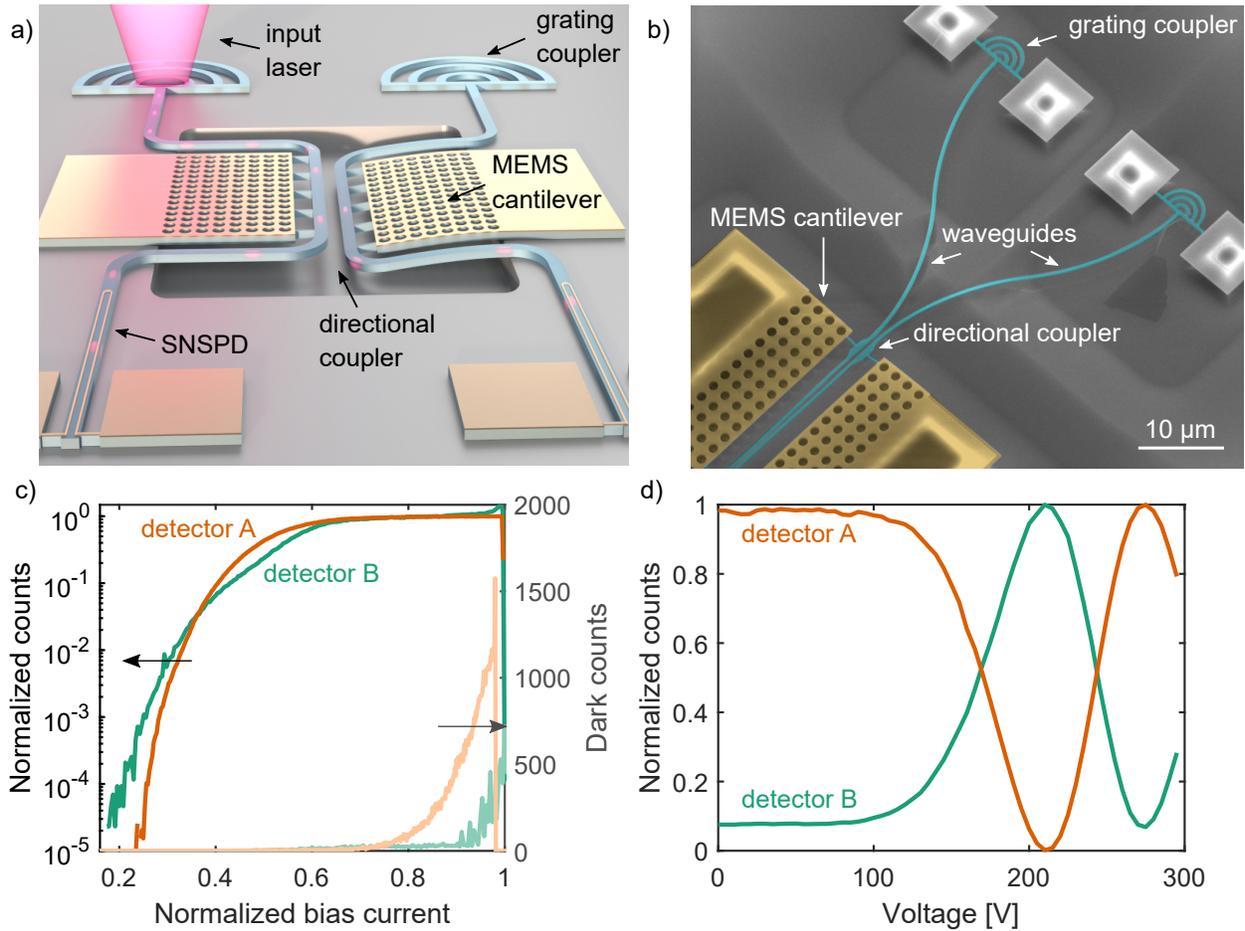

Figure 1: a) Artist view of the demonstrated device, composed of grating couplers for light input and a MEMS reconfigurable beam splitter connected to two superconducting single–photon detectors. b) False–colored SEM of the input section of our device, showing the waveguides and grating couplers, and the MEMS actuator and electrodes. c) Photon count rate at a wavelength of 795 nm for the two on–chip detectors and the corresponding dark counts. d) Normalized measured photon detection counts versus MEMS actuation voltage.

## Low–power microelectromechanics with superconducting detectors

Here, we demonstrate capacitive MEMS tuning as an SNSPD–compatible reconfiguration mechanism suitable for large–scale quantum PICs. Figure 1a shows a schematic of our device, and Fig. 1b shows a scanning electron microscope (SEM) image highlighting the two grating couplers for coupling of input light and part of the MEMS–tunable beam splitter. SEM images of the other parts of the device, including the two output SNSPDs can be found in the Supplementary.

We use the same NbTiN layer to build the MEMS actuators, electrical connections, contact pads, and single–photon detectors. Our fabrication process, described in the Supplementary, is largely enabled by the etch–resistance of NbTiN to hydrofluoric acid. The MEMS actuator consists of a NbTiN-on-$Si_3N_4$ cantilever, suspended over the Si substrate. The application of an electrical potential between the cantilever and the substrate forms a charged capacitor, which is subject to an attractive force that bends the cantilever vertically. The cantilever is attached to one of the two air–clad waveguides forming a directional coupler, and actuation results in an increase of the vertical separation between the waveguides, as illustrated in Fig. 1a. This, in turn, reduces the modal overlap, and thus changes the splitting ratio of the beam splitter, which we measure using the integrated SNSPDs. The measured tuning curve is shown in Fig. 1d, and follows our simulations (see Supplementary), yielding a high extinction ratio between detectors of $28.1\,\text{dB}$, and on–off ratio for individual ports up to $27.5\,\text{dB}$.

The normalized frequency response of our device is shown in Fig. 2a. We observe constant modulation amplitude up until the first mechanical resonance frequency between $1\,\text{MHz}$ to $2\,\text{MHz}$, in line with our simulated value of $1.6\,\text{MHz}$. Our device presents stable and reliable operation, with hysteresis below $2.4\,\%$ and power stability with a standard deviation below 0.5 % over 60 min. During the frequency sweeps, we performed more than 20 million switching events proving the durability of integrated MEMS devices. To further demonstrate device robustness, we performed three cool-down cycles and confirmed the operation of the MEMS and detectors. In terms of power consumption, for a 100 kHz drive and full splitting ratio tuning, we estimate a power consumption below 75 µW, dominated by the capacitance of our contact (the MEMS actuator consumes 2.24 µW under these conditions). This power, only consumed during dynamic actuation, is dissipated along the non–superconducting transmission line, which is limited to the off–chip components, far from our SNSPDs. Due to the high insulation of vacuum and $SiO_2$ in our capacitor, leakage currents are minimal, which leads to power dissipation in the fW–range (see Supplementary) despite high DC voltage actuation. See Supplementary for more information on frequency response, hysteresis, device stability, and power consumption.

## An on–chip power meter with high dynamic range

Although measuring the power in any optical setup is a mundane task, it is crucial for the setup, troubleshooting, and success of any measurement. While a macroscopic power meter can be found in most beam paths, PICs make access to the optical signal more challenging and require integrated photodetectors. In PICs, these are commonly built using complex processes such as heterogeneous integration of Ge or InGaAs on Si.[26] In contrast, we combine the two SNSPDs and the reconfigurable MEMS, fabricated with our simple fabrication process and using the same NbTiN layer, to demonstrate a power meter with a linear dynamic range exceeding 90 dB and sensitivity down to the single–photon level.

Figure 2c shows three switchable regions of our high dynamic range power sensor that are connected by measuring detection counts while sweeping the MEMS actuation voltage. The first range is measured with most of the optical power routed into detector A. The second range is covered by detector A and detector B, where most of the optical power is routed to the lower efficiency detector B. The third range is covered by detector B with most of the optical power routed to detector A, thereby saturating detector A. The low efficiency of detector B, which enables this measurement, is attributed to fabrication variations and could be carefully engineered by varying the nanowire length coupled to the waveguide mode.[27]

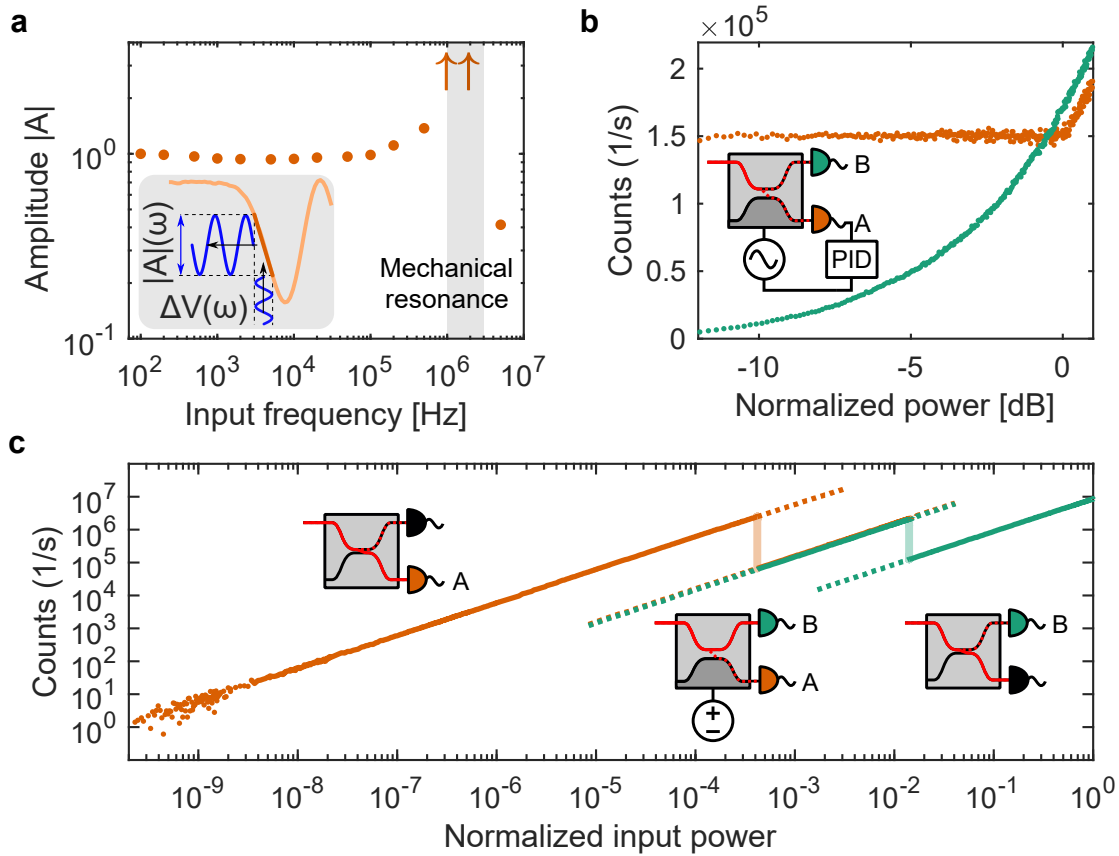

Figure 2: a) Frequency response of the tunable beam splitter measured using an on-chip detector up to the first resonance frequency normalized to the DC amplitude. The arrows represent the low bound of those measured amplitudes. Inset: this measurement was performed by translating the actuation voltage into a modulation of the splitting ratio. b) PID–controlled power stabilization on one arm (detector A, orange) using the on–chip detectors and the MEMS tunable beam splitter. The other arm (detector B, green) detects the rerouted power. c) 90 dB dynamic range on–chip photodetector that combines a high and a low–efficiency detector with switchable measurement ranges. The vertical connections are the measured counts while changing the MEMS voltage from 0 V to 196.5 V.

Our measurement range is in line with dedicated commercial devices and experimental demonstrations[28,29] but is unique in that it is sensible at the single–photon regime, as we demonstrate in following sections. This high sensitivity to very low power levels makes our sensor ideal for tap–based circuit sensing where only a fraction of the light in the waveguide is available to the photodetector. This is crucial for built–in self–tests and wafer–level testing.[30] In the following section, we use this on–chip power sensor to generate direct feedback for power control in our circuit.

## On–chip power stabilization

The large tunable extinction ratios and near–MHz speed of our device, combined with the on–chip detectors, enable dynamic stabilization of optical power in one of the beam splitter arms. Power stabilization and control is ubiquitous in classical and quantum optics experiments,[31] with key examples being spectroscopy[32] and deterministic single-photon generation.[33] In current experiments, this is implemented off–chip and requires support hardware and laboratory space, which can be minimized or eliminated by using our on–chip device. While on–chip experiments present higher stability within the chip, the input coupling efficiency of an off–chip excitation laser is highly sensitive to polarization and mechanical movement. In a packaged photonic circuit with on-chip light source not only thermal fluctuations but also fabrication process variations and aging effects need to be compensated.[34] We address these problems by providing direct feedback on the MEMS beam–splitter using the power measured on the on–chip detector A, while the excess power is routed into the second arm (detector B). Figure 2b shows the measurement results and a schematic of our device. A PID feedback loop uses the detection counts to stabilize the measured power without manual intervention. The circuit yields stable (within $1.3\,\%$ standard deviation) on–chip power while we tune the off–chip input power over 12 dB. The control loop was run every 100 ms and the input laser power was swept from $0\,\mu W$ to $300\,\mu W$ in $1\,\mu W\,s^{-1}$ steps.

## On–chip reconfigurable single–photon detection

To demonstrate the performance of our device in the single–photon regime for quantum PICs, we characterized it using single photons from an on–demand single–photon source. The source consists of a GaAs quantum dot excited via two–photon resonant excitation[35] with a 320 MHz repetition rate and a pulse length of 40 ps. Single photons from the exciton transition with an energy of 1.5636 eV (795 nm, spectrum in the inset in Fig. 3a), are coupled via an optical fiber and free–space optics into our device using one of the grating couplers.

We performed a lifetime measurement on–chip, as seen in Fig. 3a, resulting in an exciton lifetime $\tau_X = 232 \pm 2\,\text{ps}$, in line with the off–chip measurement of $\tau_X = 216.3 \pm 0.6\,\text{ps}$. We then measured single–photon purity by performing a Hanbury Brown and Twiss (HBT) measurement by routing the photons from our source into an off-chip beam splitter with one arm coupled through our device into the on–chip detector A, and the other arm going into a commercial off–chip SNSPD. Figure 3b shows our measurement, yielding a $g^{(2)}(0) = 0.11 \pm 0.05$, clearly showing anti–bunching and single–photon detection, in line with the $g^{(2)}(0) = 2.97 \cdot 10^{-4}$ measured using two commercial SNSPDs with the same emitter shown in Fig. 3d.

The discrepancy between our on–chip and off–chip $g^{(2)}(0)$ is due to the limited optical coupling efficiency into the chip leading to a degraded signal to noise ratio between dark counts of the detector and detection events due to single photons. This can be improved by low–loss coupling between quantum emitters and the PIC, either via low–loss fiber coupling,[36] or quantum emitter integration using monolithic[37] or hybrid approaches.[17]

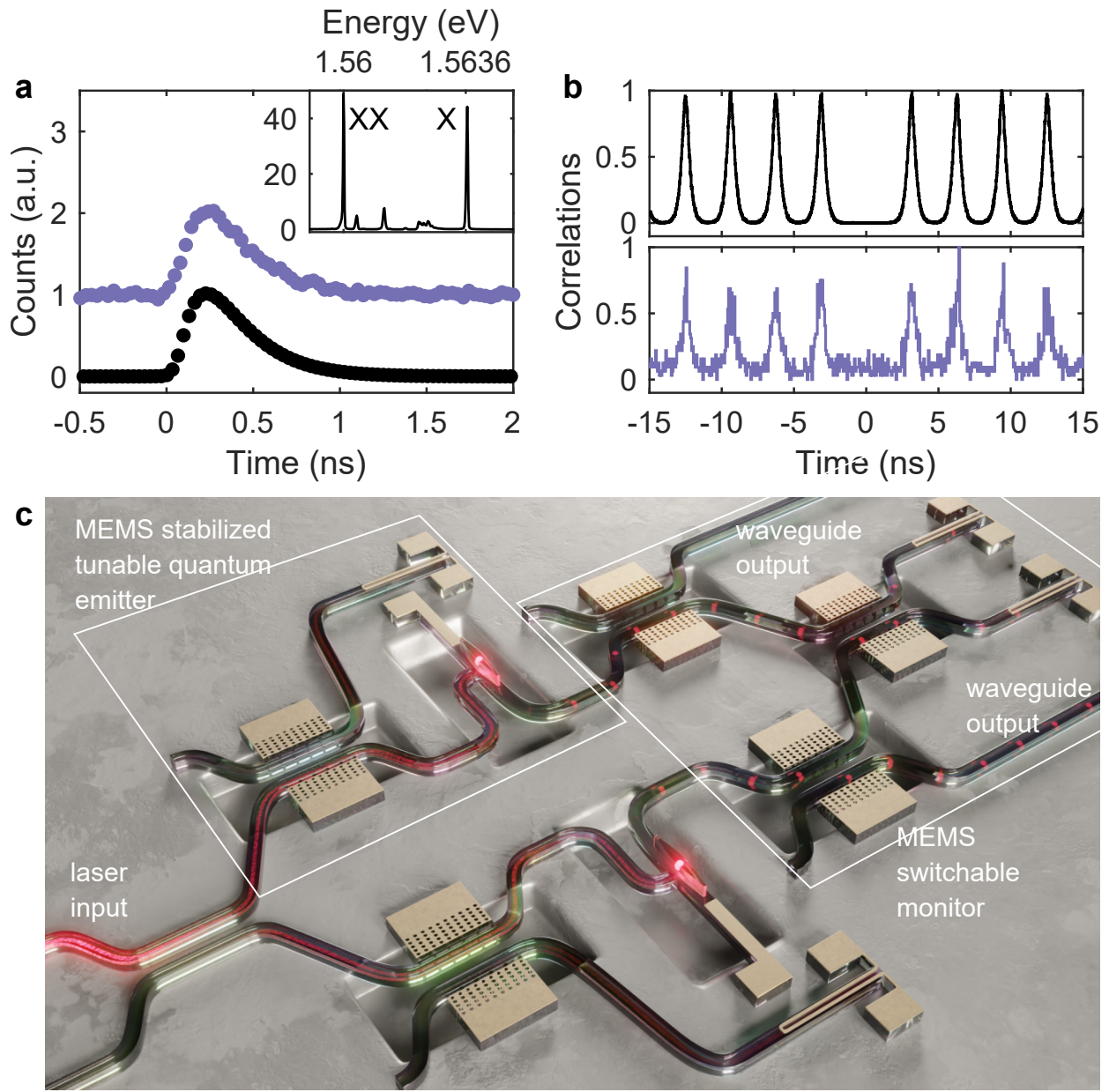

Figure 3: a) Lifetime measurements using our device (purple) compared to commercial fiber–coupled SNSPDs (black). Inset: spectrum of the deterministically excited quantum dot used in this work, under $\pi$ pulse excitation, with highlighted exciton (X) and biexciton (XX) lines. b) HBT measurements, showing a comparison of detection off–chip (top) and detection with one SNSPD on–chip (bottom). c) Example of a monitored and stabilized on–chip photon source as a near–term application of our technology. The device includes two power stabilizers connected to two MEMS tunable quantum emitters and MEMS splitters for switching into an HBT/HOM monitoring circuit.

# Discussion

Our proof-of-principle implementation of this technology could benefit from established fabrication processes and reach state-of-the-art performance. For example, improved SNSPD designs and readout electronics have demonstrated longer wavelength detection,[20] shorter reset times,[38] and sub-3 ps detection jitter.[39] In an industrial fabrication process, fabrication yield is expected to be no limiting factor.[40] For high frequency and high voltage signals ($\approx$150 V and >10 kHz), our MEMS actuation signal creates detection events on the detector channels even on unbiased detectors due to the large amplification in the readout channel (see Supplementary for a more in–depth discussion). This can be improved by better electrode design on–chip and by reducing the actuation voltage.

MEMS technology is inherently scalable and reliable, as evidenced by the success of MEMS sensors in commercial electronics. For PICs, capacitive MEMS have demonstrated low–voltage actuation and excellent scalability,[16,41,42] which makes it likely that foundries will offer photonic MEMS soon, and our platform provides a clear pathway towards simultaneous integration of MEMS and SNSPDs in a simple fabrication process. Cryogenically–compatible MEMS in quantum PICs provide seamless integration of displacement and strain actuators on–chip. This has been shown to be a fundamental part of other PIC building blocks, such as tunable filters[43] and phase shifters.[44,45] Additionally, MEMS have demonstrated compatibility with integrated single–photon emitters for routing[46] and filtering,[47,48] and the strain distribution created by capacitive MEMS has been used to improve the spectral overlap of quantum emitters,[49,50] and the spin coherence of color centers.[51] The presented platform can be leveraged for these applications, as shown in Fig. 3c, in which the quantum emitters are strain–tuned into spectral alignment using a MEMS cantilever.

Applications of the presented device span many relevant quantum optics experiments. The analog tunability of our device enables fine optimization and stabilization of beam splitting ratios, critical for random number generation, and accurate HBT and Hong-Ou-Mandel (HOM) measurements. Fast reconfiguration in the MHz regime can be used to tap and measure properties of quantum emitters on–chip. This can be readily applied to existing quantum communication protocols such as quantum key distribution to identify multiphoton or blinding attacks. Additionally, by changing the device geometry to mismatch the waveguide modes (i.e. reducing the width of one of the waveguides), the device can act as a phase shifter,[43] which, together with passive beam splitters, form a complete set of components for arbitrary linear optics and linear optical quantum computation.[8] Figure 3c shows a monitored and stabilized on–chip single–photon source as a near–term application of our technology. The device consists of three parts: stabilized laser excitation of tunable single–photon emitters, switching between monitors and output, and monitoring circuit. A laser input is sent through a passive ∼50:50 beam splitter into two $\pi$–pulse power stabilizers each formed by a MEMS splitter and an SNSPD. The laser pulses then excite two MEMS–tunable single–photon sources, and their emission is routed into MEMS beam splitters which act as a switch between the output waveguides and the integrated monitoring. The on–chip monitoring circuit is based on a single MEMS beam splitter terminated with two SNSPDs. This enables switching between the analysis of three crucial aspects of a single–photon source: i) emitter intensity by routing all the emitted photons into an SNSPD (MEMS splitter in 100:0 splitting), ii) photon antibunching from each emitter (MEMS splitter in 50:50) by routing only the photons from one emitter into the device forming an HBT setup, and iii) photon indistinguishability (MEMS splitter in 50:50) by routing the photons from both emitters into quantum interference forming a HOM setup. The device can thus discretely or continuously monitor emitter properties by switching from on–chip to off-chip configurations or by using the MEMS splitters as a tap.

## Conclusion

We have demonstrated the on–chip compatibility of MEMS reconfigurable photonics with superconducting single-photon detectors and used it to develop three key elements of quan-tum optics experiments. We measured routing of classical and quantum light with a high extinction ratio, a photodetector with high dynamic range, and input power stabilization. Using our device, we performed on–chip single–photon measurements from a quantum dot source, yielding a lifetime of 232 ps and $g^{(2)}(0) = 0.11$. Our results show that the combination of MEMS and SNSPDs enables the on–chip integration of not only the main building blocks of quantum optics, but also devices for adaptive control, monitoring, and stabilization of classical and quantum optics. The presented technology can overcome current roadblocks towards large–scale quantum optics, and foster applications in quantum communication, metrology, computing, and simulation.

# Acknowledgement

The authors thank Dr. Umer Shah and Prof. Joachim Oberhammer for the amplifier, and Aleksandr Krivovitca for his help with the critical point dryer. This project has received funding from the European Union's Horizon 2020 research and innovation program under grant agreement No. 820423 (S2QUIP). S.G. acknowledges funding from the Swedish Research Council under Grant Agreement No. 2016-06122 (Optical Quantum Sensing). A.R. acknowledges the support of the LIT Secure and Correct Systems Lab, financially supported by the State of Upper Austria, and the Austrian Science Fund (FWF): P29603, I4320. V.Z. acknowledges funding by the Knut and Alice Wallenberg Foundation (KAW, "Quantum sensors") and the Swedish Research Council (VR, grant No. 638-2013-7152 and grant No. 2018-04251). C.E. acknowledges funding from the Swedish Research Council (2019-00684). C.E. is currently with the Quantum Photonics Laboratory, Massachusetts Institute of Technology.

# Author contributions

C.E. conceived the device, and C.E., S.G., V.Z. and K.D.J. conceived the demonstrator experiments. C.E. performed the design and simulations. C.E., S.G., J.Z., A.W.E, S.S. developed and performed device fabrication. S.F.C. and A.R. developed and performed the quantum dot growth and characterization. S.G., C.E., L.S., and K.D.J. performed the measurements. C.E. and S.G. wrote the manuscript, and all authors revised the manuscript. C.E., V.Z., and K.D.J. supervised the project.

# Competing interests

The authors declare no competing financial interests.

# Supplementary

## Sample description

Scanning electron micrographs (SEMs) of our device can be found in Fig. 1b and Fig. 4. The waveguides are 250 nm thick and 500 nm wide, with $SiO_2$ bottom cladding and air top cladding, adiabatically coupling into 400 nm wide, air–clad waveguides in the directional coupler region. The waveguides bend at a radius of 25 µm, which was designed conservatively to avoid optical radiation loss. The grating couplers are air–clad, and have a period of 900 nm and a duty cycle of 50 %, and were designed for objective coupling following Zhu et al.[52], and optimized in previous fabrication runs by sweeping and measuring grating transmission, which we measured to be around 10 % for a microscope objective with a NA of 0.65. Near the SNSPD, the waveguide width tapers to 550 nm. The SNSPD is fabricated in a 9 nm thin NbTiN film, and features a hairpin design with 65 nm width and 40 µm total length. The MEMS actuator is a cantilever of length 9.5 µm and width (i.e. directional coupler length) 80 µm, with 1 µm diameter holes for hydrofluoric acid wet under–etching. The anchors to connect the suspended waveguides to the rest of the chip widen to 650 nm with a 1.5 μm long taper, followed by a 1.5 μm straight section before tapering down symmetrically. The 6 clamps connecting the MEMS actuator to the tapered waveguide are 300 nm wide and 1.6 μm long, and are separated with a pitch of 23.5 μm. The directional coupler was designed with symmetrical cantilevers, with only one cantilever electrically contacted to act as the MEMS actuator.

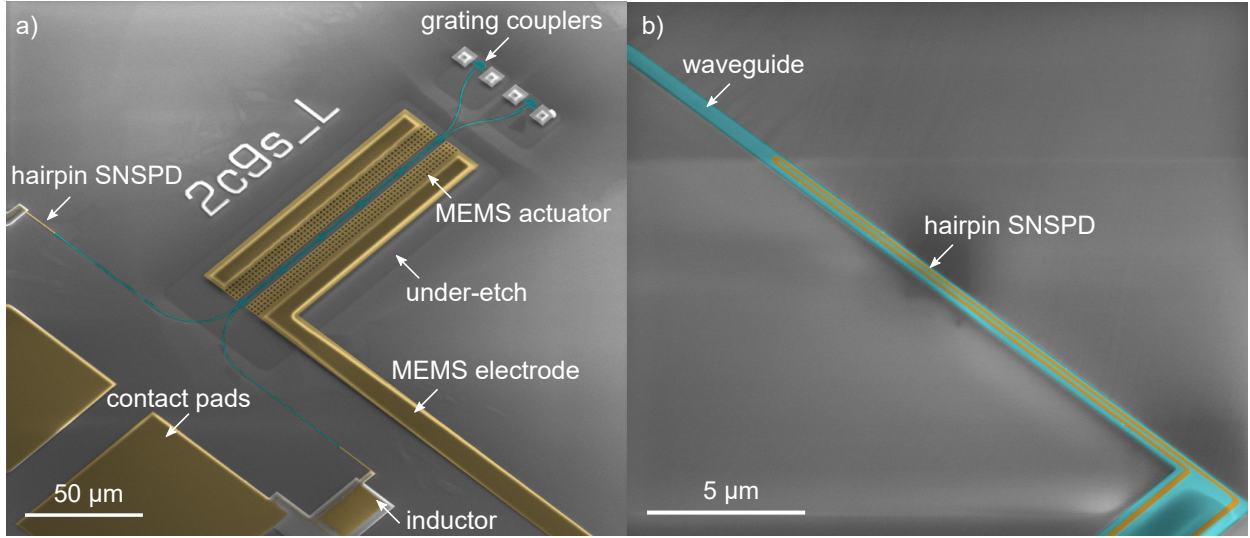

Figure 4: SEM images of our device. a) Full view including the two input grating couplers, the suspended directional coupler and routing waveguides (blue) and MEMS actuators, electrodes, SNSPDs and contact pads (gold). b) Close–up of one of the waveguide–coupled SNSPDs.

## Sample fabrication

The sample fabrication started with a 250 nm thin film of stoichiometric $Si_3N_4$ on 3.3 μm $SiO_2$ on a silicon handle substrate, provided by a foundry (Rogue Valley Microdevices). We deposited a 9 nm film of $Nb_{0.86}Ti_{0.14}N$ by reactive co-sputtering from separate Nb (200W, DC) and Ti (200W, RF) targets at room temperature (Fig. 5a) in nitrogen and argon atmosphere.[21] After Cr/Au marker lift-off (Fig. 5b), aligned electron beam lithography followed by $CF_4$-based reactive ion etching resulted in patterned NbTiN nanowires, contacts, and MEMS electrodes (Fig. 5c). A second electron-beam lithography and a $CHF_3$-based reactive ion etching yielded the $Si_3N_4$ waveguide devices (Fig. 5d). Then, we used optical lithography with positive–tone resist to open windows for a wet buffered hydrofluoric acid (BHF) process under–etching the $SiO_2$ film under the MEMS actuators and grating couplers. To avoid the collapse of the suspended actuators due to capillary forces, we dried the sample using critical point drying (CPD, Fig. 5e).

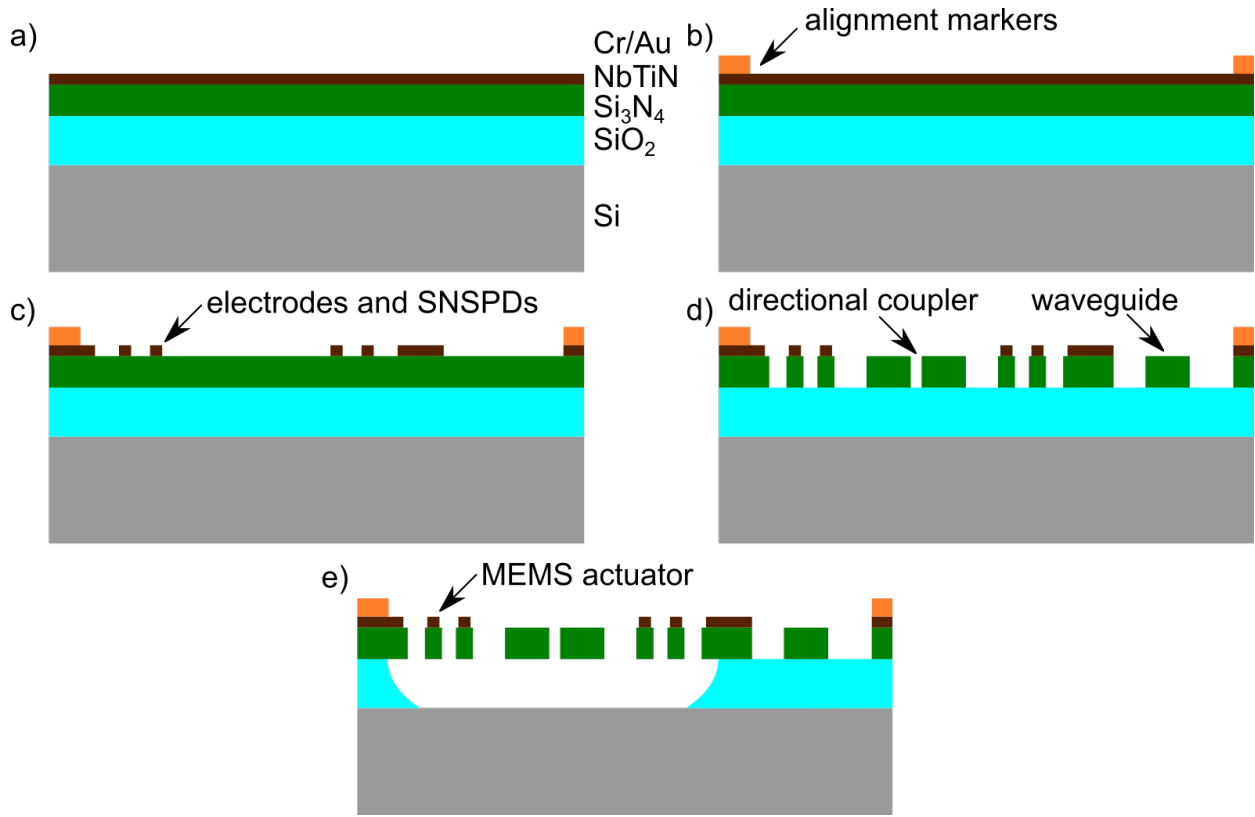

Figure 5: Cross-section schematic of the fabrication process for the MEMS reconfigurable device with SNSPDs. a) Deposition of NbTiN on a foundry–grown $Si_3N_4/SiO_2/Si$ substrate, followed by b) Au/Cr lift–off of markers, c) NbTiN and d) $Si_3N_4$ patterning using electron beam and reactive ion etch, and e) BHF under–etching through resist windows patterned via photolithography followed by CPD.

## MEMS design and simulations

We performed eigenmode simulations of the directional coupler supermodes using COMSOL Multiphysics at 795 nm wavelength of two $Si_3N_4$ ($n_{Si_3N_4} = 2$) coupled waveguides (width 400 nm, height 250 nm, horizontal separation 200 nm) clad by air ($n_{air} = 1$).

From coupled-mode theory, the power $T$ at one of the output ports of a directional coupler ($n_{eff,i}$ is the effective index of the mode in the respective arm and $L_{wg}$ is the length of the coupling region)

$$T_1 = sin^2\left(\frac{(n_{eff,s1}(y) - n_{eff,s2}(y))\pi L_{wg}}{\lambda}\right), \qquad (1)$$

with $T_2 = 1 - T_1$ in the other port

For the mechanical simulations, we assumed no initial stress with thickness 245 nm (combined $Si_3N_4$ and NbTiN films, assuming a minor etch of the $Si_3N_4$ due to finite selectivity of the HF underetch), and a Young's modulus of 180 GPa (reported $Si_3N_4$ values range from 120 GPa to 250 GPa). Initial capacitor separation was 3.535 µm (air bottom cladding plus $Si_3N_4$ thickness), and vacuum permittivity between the electrodes ($\epsilon = \epsilon_0$).

Using Hooke's law, the restoring force for an ideal spring is $F_k = ky$, with $y$ vertical displacement. To calculate the spring constant $k$, our mechanical simulations follow the Euler-Bernoulli beam theory for a clamped-free beam (rectangular cross-section of width $w$, thickness $t$, length $L$, and for a material with Young's modulus $E$), with $k = \frac{3EI}{L^3}$, and its moment of inertia with respect to its neutral axis is $I = \frac{wt^3}{12}$, leading to

$$F_k = \frac{Ewt^3}{L^3}y. \qquad (2)$$

This can be used to calculate the first resonance frequency as $f_r = \frac{1}{2\pi}\sqrt{\frac{k}{m}}$, with $m$ the cantilever mass. Our calculations yield $f_r = 1.6$ MHz, in line with our measured value.

For the electrostatic force, we assume a parallel-plate capacitor of area $wL$. Assuming negligible fringing fields, the electrostatic force is

$$F_e = \frac{\varepsilon w L V^2}{2(y_0 - y)^2}. \qquad (3)$$

The force balance combining Eqs. 2 and 3 yields a relation between actuation voltage

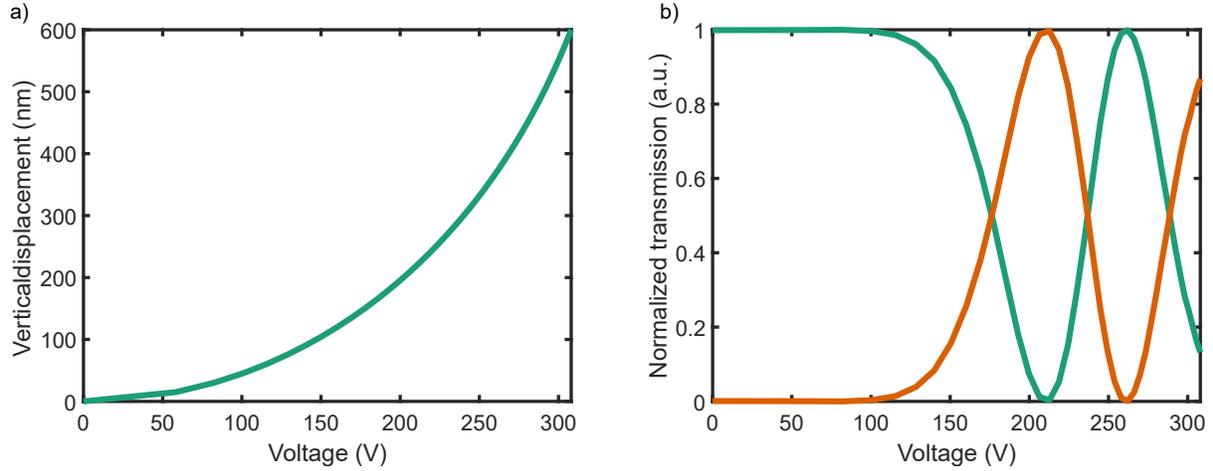

Figure 6: Simulation results for our device. a) MEMS vertical displacement versus actuation voltage. b) MEMS beam splitter splitting ratio versus actuation voltage.

and displacement.

$$V^2 = \frac{2Et^3 y(y_0 - y)^2}{4\varepsilon L^4}. \tag{4}$$

Figure 6a shows our simulated vertical displacement under MEMS voltage actuation.

Equation 4 can be solved analytically using Cardano's cubic solution, although for simplicity we solved it here numerically. Inserting the solution into Eq. 1 we obtain a relation between the power transmission and the actuation voltage $T_1(V)$, which we plot in Fig. 6b using our design parameters.

## Power consumption estimate

Electrical energy ($E_c$) from charging and discharging a capacitor with capacitance $C$ is $E_c = CV^2$, which for a driving frequency of $f$ translates into a power of $P = fCV^2$. Compared to charge–discharge power, the contribution of leakage currents in such large capacitors is negligible. The capacitance of our device can be estimated from a parallel–plate capacitor with its area dominated by the $6400\,\mu m^2$ contact pad and a $SiO_2$ ($\varepsilon = 3.9$) gap of $3\,\mu m$. For an AC voltage actuation of $100\,V$ on top of a DC bias of $100\,V$ to cover the full splitting ratio range, and an actuation frequency of $100\,kHz$ (right below the resonance), the power consumption amounts to less than 75 μW. We note that for lower–voltage MEMS actuator design, such as those reviewed in Ref.,[16] the power consumption is usually in the nW range due to a capacitance not dominated by the contact.

The power consumption due to leakage current can be estimated using the resistance of the dielectric between the parallel plates as $P_{leak} = V^2 R$. The high resistivity of SiO$_2$ and of vacuum leads to a negligible leakage currents, and thus to power dissipation below 10 fW. In a foundry device, however, the leakage current will be likely dominated by the control electronics and the routing of electrical lines.

## Experimental setup

The sample was silver–glued to a custom–designed printed circuit board (PCB), and wire-bonded using an Al wedge bonder. The PCB was then mounted in a dilution cryostat (Bluefors) with optical window access and a sample stage temperature below 100 mK. The PCB was connected to coaxial cables leading to room-temperature, where the SNSPDs are biased and amplified using a commercial driver system (Single Quantum Atlas). The MEMS components are either directly driven using a high–voltage power supply (Keithley 2410) or through a function generator amplified using a high–speed high–voltage amplifier (Falco Systems, WMA-300). Light was coupled into the chip through an objective (50x, NA 0.82, Partec) inside the cryostat from a CW-laser at 795 nm wavelength, a pulsed laser (795 nm, 2 ps) or the quantum dot single–photon source. The polarization was controlled using a $\lambda/2$ - waveplate on the input path. The coupling efficiency is estimated at $-9.3$ dB by measuring the input and out–coupled power of a U-turn reference waveguide.

## Single–photon measurements

The quantum dot sample was grown at Johannes Kepler University Linz via molecular beam epitaxy by etching nanoholes into an AlGaAs layer before filling the holes with GaAs.[53] The quantum dot is embedded in a $\lambda$–cavity with a ditributed Bragg reflector consisting of 9 thin film pairs (GaAs/AlGaAs) below and 2 pairs above the cavity. A solid immersion lens was placed on top of the sample to increase light extraction efficiency further. The sample was placed inside a closed–cycle Helium–cryostat operating at 5 K and emitted light was collected and collimated by an aspheric lens inside the cryostat with a numerical aperture of about 0.5. The light was coupled into the collection path via a 90:10 beam splitter in order to minimize losses and the excitation laser was filtered out with three tunable notch filters with a bandwidth of 0.4 nm. An optical fiber was used both as a pinhole to further suppress the excitation laser and to route the light to a transmission spectrometer. Here, biexciton and exciton emission were coupled into separate optical fibers.

The exciton emission was split at 50:50 fiber beam splitter and both arms were connected to two superconducting nanowire single–photon detectors (Single Quantum EOS) with efficiencies of 50 % and 60 %, timing jitters of 20 ps and 30 ps, and dark count rates of 0.006 cts/s and 0.017 cts/s. In the measurement with an on–chip detector, one of the arms was coupled to the detectors on–chip.

## SNSPD characterization

We measured the photon count rate depending on bias current of the SNSPDs through the waveguide at 795 nm. The curves were individually normalized to the saturation plateau of our devices. The devices show a reset time given by the Gaussian decay fit of $4.73 \pm 0.03$ ns for detector A, and $4.73 \pm 0.01$ ns for detector B, shown in Fig. 7.

We measured the jitter of our integrated SNSPDs using a 2 ps pulsed laser system at 795 nm using room temperature amplification in the commercial driver system. The photodiode generated trigger signal and the SNSPD signal were correlated on a 40 GS/s, 4 GHz bandwidth oscilloscope (WaveRunner 640Zi, LeCroy). The extracted jitter from an exponentially modified Gaussian fit is 121 ps for detector A and 253 ps for detector B (see Fig. 8).

For the MEMS characterization in Fig. 1d, our absolute measured counts ranged from 4 400 246 to 7 793 for detector A, and 98 619 to 6 747 for detector B, resulting in 27.52 dB (detector A) and 11.65 dB (detector B) difference between the two extreme points. We attribute the large (44.6 times) difference in detector efficiency to film heterogeneity and fabrication variations leading to localized constrictions.[25] We use this difference in detection efficiency as a property to extend the detection range of the power sensor.

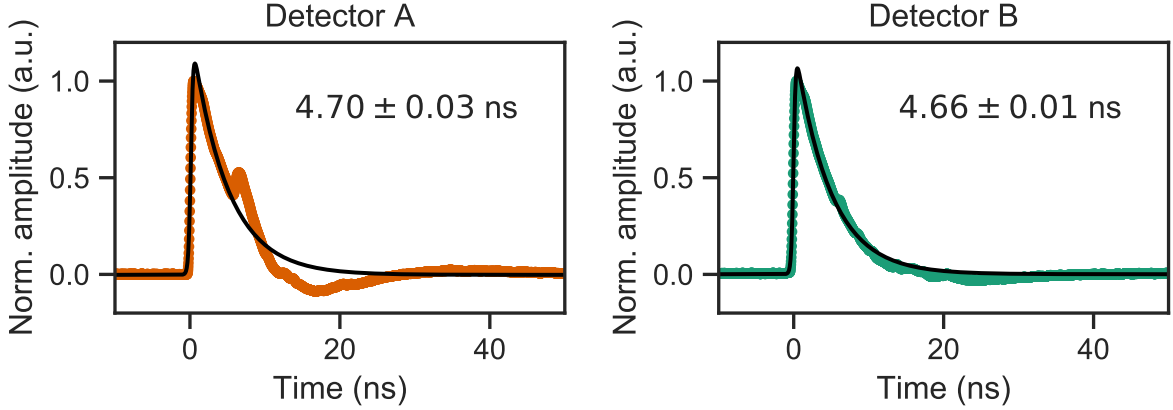

Figure 7: Averaged detection pulse of our SNSPDs with reset times given by Gaussian decay fits of $4.73 \pm 0.03$ ns (detector A) and $4.73 \pm 0.01$ ns (detector B).

## Frequency response measurement

We actuated the MEMS system using a function generator with an amplifier (amplification factor 50) as explained above. We operated the MEMS in a bias point of 165 V with an amplitude of 60 V using an in-house built high–voltage bias tee (See Fig. 9). The trigger linked to the sinusoidal excitation and the detection event of the on–chip SNSPD was connected to a time correlator (qutag, qutools). The time–tagged data was evaluated using our in-house built software, ETA.[54]

The measurements at the frequencies of 1 MHz and 2 MHz were largely amplified by the mechanical resonance, which led to a large waveform distortion (i.e. effectively actuation with amplitudes higher than our measured actuation curve in Fig. 1d) and the SNSPD response fluctuates between maximum and minimum transmission. Due to the lack of a characterization curve covering such large amplitudes, we were unable to fit an analytical model to their time response, and their amplitude in Fig. 2a is the difference between maximum and minimum transmission, and thus represents a lower bound to their amplitude. This waveform distortion, caused by a large mechanical amplification near the resonance, hampers our measurement of the exact resonance frequency and mechanical Q-factor. We expect the Q-factor to be high due to the absence of viscous damping and the low coefficients of thermal expansion at cryogenic temperatures resulting in low thermo–elastic loss.

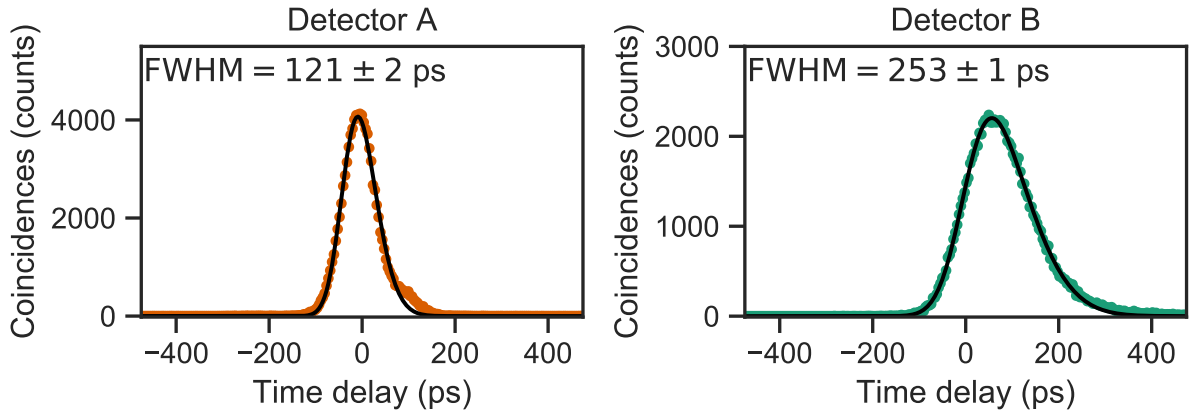

Figure 8: Jitter curves for the two presented SNSPD devices with an exponentially modified Gaussian fit with a full width at half maximum of $121 \pm 2\,\text{ps}$ (detector A) and $253 \pm 1\,\text{ps}$ (detector B).

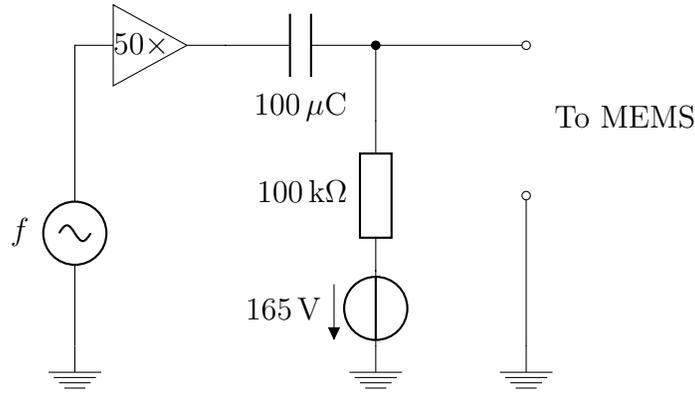

Figure 9: Circuit for the frequency measurement. The measured SSPD counts where fitted to a sine and the amplitude used as frequency response in Fig. 2a.

At frequencies above $10\,\text{kHz}$ we observe detection events on an unbiased SNSPD channel due to the large amplification on the readout chain. This is much more significant on detector B due to the low critical current, that also requires a lower trigger level. Figure 10 shows these counts for both detectors with a sinusoidal actuation between 10 kHz and 5 MHz, an amplitude of 60 V at a working point of 165 V. Through careful RF engineering on the electrodes on chip (e.g. ground planes encapsulating the signal lines) this cross-talk can be reduced or eliminated.

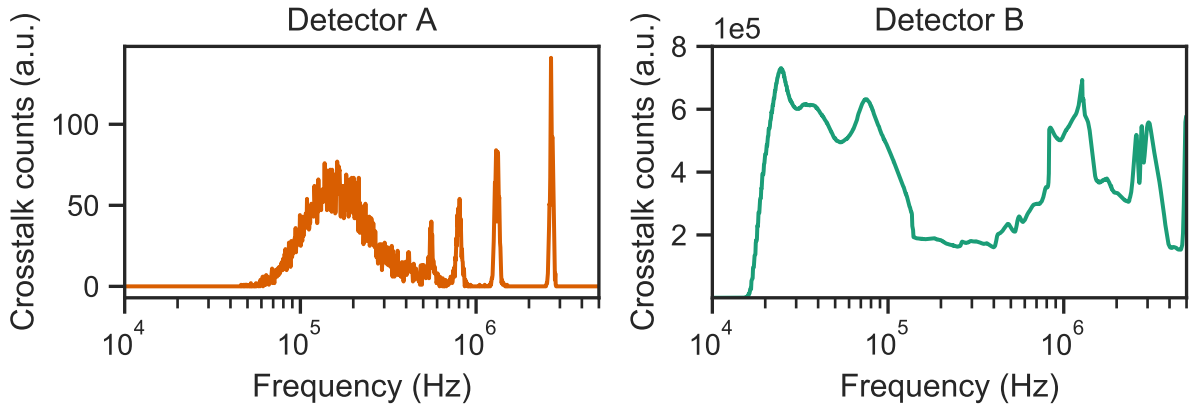

Figure 10: Detection counts generated by sinusoidal actuation (from 10 kHz to 5 MHz with an amplitude 60 V at a working point 165 V) of the MEMS on the unbiased SNSPD readout channel.

## Stability and hysteresis measurements

We measured the device stability by setting the MEMS actuator to 190 V during 60 min, and the measurement can be found in Fig. 11a. Our measurement yields constant transmission within a standard deviation below 0.5 %. The small asymmetric instability is most likely caused by setup drift (e.g. laser power or polarization fluctuations, fiber and setup vibrations) and not MEMS instability, since the cantilevers feature a high stiffness, and fluctuations would most likely happen near the MHz–range fundamental resonance.

We then performed a hysteresis measurement (Fig. 11b) by cycling the MEMS actuator voltage between 120 V and 240 V over 10 cycles at a frequency of 1 Hz, yielding a hysteresis below 2.4 % in optical transmission. This is likely an overestimated value, since this measurement will be affected by power fluctuations caused by free–space coupling and mechanical

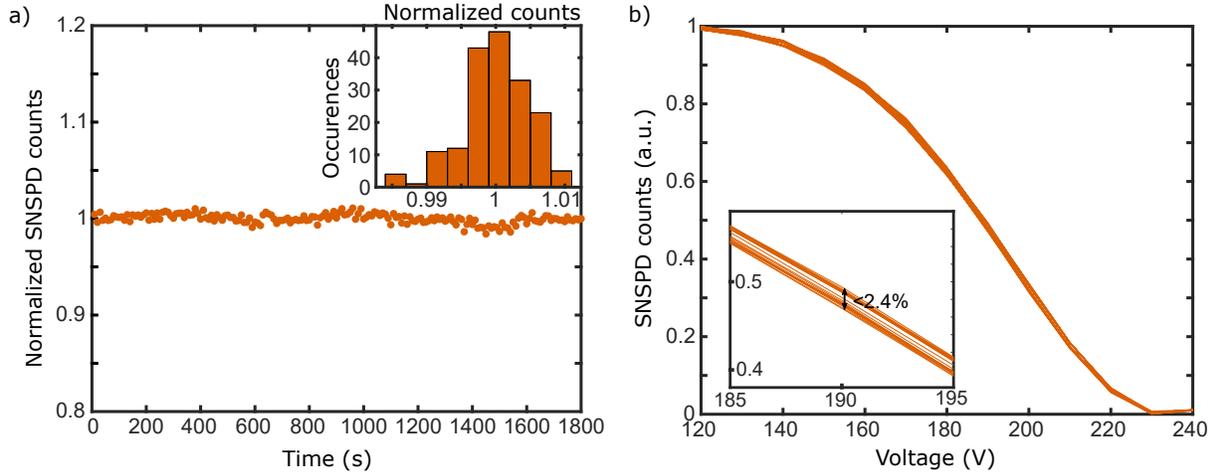

Figure 11: a) Stability measurement with our MEMS beam splitter under 190 V actuation during 60 min. Inset: distribution of the measured SNSPD counts, yielding a standard deviation below 0.5 %. b) Hysteresis measurement consisting of periodic charge–discharge cycles of the MEMS actuator between 120 V and 240 V over 10 cycles, showing hysteresis below 2.4 %.

vibrations in the measurement setup.

## Power stabilization measurements

A in–house built digital–to–analog converter (DAC) circuit and the SNSPD driver were controlled by our in–house built LabView based laboratory control system. The output of the DAC was amplified using the high–voltage high–speed amplifier as described in the previous section and fed as offset voltage to the high–voltage power supply (Keithley 2410) (See Fig. 12). The SNSPD driver was read out every 100 ms and a PID control loop stabilized the applied voltage with the photon detection events as the process variable. We tested the control loop by ramping the input power from $0\,\mu W$ to $300\,\mu W$ with $1\,\mu W\,s^{-1}$.

## High–dynamic range detector

We illuminated our device using a continuous wave laser at 795 nm and swept the input power using a fiber–coupled attenuator while recording the SNSPD counts per second. To reach the wide range of input power we additionally added absorption based attenuators ($-130\,dB$, $-120\,\mathrm{dB}$, $-70\,\mathrm{dB}$, $-40\,\mathrm{dB}$, $-10\,\mathrm{dB}$) into the free–space path. The different input power regions per leg are aligned by matching the slope of the measured detection counts based on a linear fit. We connected the measurements at the switching points of our high–dynamic range detector by keeping the input power constant while ramping the MEMS actuation voltage between $0\,\mathrm{V}$ and $196.5\,\mathrm{V}$. The power axis is normalized to the maximum measurable power by the power sensor.

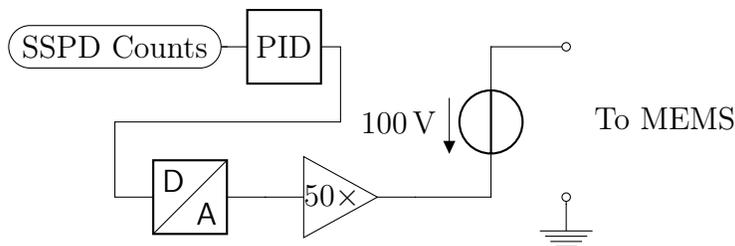

Figure 12: Circuit for feedback loop stabilizing the count rate measured on the SSPD. The PID controller was implemented in LabView.